\documentclass[10 pt]{article}
\usepackage{amsmath,amsfonts,amssymb,bm}
\usepackage{float}
\usepackage{subfigure}
\usepackage{graphicx}
\usepackage{wrapfig}
\usepackage[margin=1in]{geometry}
\usepackage{hyperref}

\begin{document}
\title{The Electromagnetic Field as a Synchrony Gauge Field}
\author{Robert D. Bock\footnote{robert at r-dex.com} \\ R-DEX Systems \\ \url{www.r-dex.com}}
\date{\today}

\maketitle

\begin{abstract}
\noindent Building on our previous work, we investigate the identification of the electromagnetic field as a local gauge field of a restricted group of synchrony transformations.  We begin by arguing that the inability to measure the one-way speed of light independent of a synchronization scheme necessitates that physical laws must be reformulated without distant simultaneity.  As a result, we are forced to introduce a new operational definition of time which leads to a fundamental space-time invariance principle that is related to a subset of the synchrony group.  We identify the gauge field associated with this new invariance principle with the electromagnetic field.  Consequently, the electromagnetic field acquires a space-time interpretation, as suggested in our previous work.  In addition, we investigate the static, spherically symmetric solution of the resulting field equations.  Also, we discuss implications of the present work for understanding the tension between classical and quantum theory.
\end{abstract}

\noindent \textbf{KEY WORDS}:  relativity, electromagnetism, time, simultaneity, Maxwell's equations, gauge theory, quantum theory   
                             
\section{\label{sec:introduction}Introduction}
Previously, we developed a gauge theory of the combined gravitational-electromagnetic field by expanding the Poincar\'e group to include clock synchronization transformations \cite{bock2015a}.  We showed that electromagnetism can be interpreted as a local gauge theory of the synchrony group.  The resulting field equations for the electromagnetic field acquired nonlinear terms and electromagnetic gauge transformations acquired a space-time interpretation as local synchrony transformations.  However, the identification of the electromagnetic field as a synchrony gauge field encountered an obstacle, given that the synchrony group produces three fields that satisfy Maxwell's equations to lowest order rather than a single U(1) field.  This suggests that the additional gauge fields represent new fields unrelated to the electromagnetic field or the full synchrony group does not represent a fundamental symmetry of nature.  In our previous paper, we proposed that nature hides this triplication via symmetry or by restricting the fundamental invariance group to a subset of the full synchrony group. 

The objective of this paper is to explore further the identification of the electromagnetic field as a local gauge field of a restricted group of synchrony transformations in order to resolve the dilemma we encountered previously.  In the following, we demonstrate that the fundamental invariance principle that leads to the electromagnetic field via the gauge prescription is indeed related to the synchrony group.  This invariance principle is uncovered by introducing a new operational definition of time based on the inability to define distant simultaneity.  By identifying the gauge field associated with this new invariance principle with the electromagnetic field, we provide a space-time interpretation of the electromagnetic field.  The static, spherically symmetric solution is also discussed.  We conclude by exploring the implications of the present investigation for resolving the tension between classical and quantum theory.

\section{\label{sec:reformulation}The Reformulation of Physical Laws Without Distant Simultaneity}
Einstein revolutionized our understanding of the nature of time with his theory of special relativity \cite{einstein1905}.  Breaking with his predecessors who identified a single global time for all observers irrespective of their relative motion, Einstein asserted that each inertial frame of reference carries its own temporal coordinate.  As is well known, special relativity has enjoyed overwhelming success and has had a profound effect on the subsequent development of physics.  

Fundamental to Einstein's formulation of special relativity is the definition of distant simultaneity in a single inertial frame of reference.  As is well known, Einstein used light propagation to define simultaneity for spatially separated clocks.  He proposed a seemingly consistent definition of distant simultaneity by assuming the one-way speed of light is identical to the two-way speed of light, although he recognized that this assumption was a matter of convention.  This topic has since received significant attention in the literature and is known as the \textit{conventionality of simultaneity}.  Whereas supporters of the conventionality of simultaneity (e.g., \cite{reichenbach1958, grunbaum1973, zhang1997,anderson1998}) argue that clock simultaneity is an arbitrary convention that permits different one-way speeds of light but preserve the experimentally measured two-way speed of light, opponents of the conventionality of simultaneity (e.g., \cite{salmon1977, malament1977, ohanian2004}) argue that standard synchrony defined by Einstein synchronization is the only clock synchronization convention that is permitted by fundamental physical laws and that the one-way speed of light is equal to the experimentally measured two-way speed of light. 

The inextricable link between the one-way speed of light and the synchronization of distant clocks combined with the failure to experimentally measure the one-way speed of light independent of a synchronization scheme strongly supports the thesis of the conventionality of simultaneity.  However, even if one accepts the conventionality of simultaneity thesis, one must admit that distant simultaneity itself has a precarious ontological status in physics.  If there is no way to measure the one-way speed of light independent of a synchronization scheme, then we must recognize that distant clock synchronization is not permitted by nature and is introduced only for the expedience of calculations.  Whereas the two-way speed of light is a real physical quantity, we must not assume the one-way speed of light can be similarly defined.  \textit{Therefore, moving beyond the thesis of the conventionality of simultaneity, we propose to reformulate physical laws without distant simultaneity}.

To be clear, we are not claiming that special relativity is wrong; there is no denying the positive impact Einstein's theory has had on modern physics.  We are claiming that special relativity employs distant simultaneity only to simplify the mathematical formalism and in so doing has tacitly introduced unmeasurable and hence unphysical quantities into our description of space-time.  While this formalism has proven to be powerful, we must examine it carefully to ensure we are not propagating false assumptions in the foundations of our theories.  Our experience with the concept of a one-way velocity that is defined according to simultaneous, separated clocks in our everyday lives has perhaps obfuscated its problematic ontological origin.  Therefore, we assert that our current formulation of physical laws may be a flawed, yet useful, representation of nature that must be reexamined.  

We now proceed to introduce a new operational definition of time in order to remove distant simultaneity from the formulation of physical laws.  We consider an infinite set of non-synchronized clocks situated throughout an inertial frame of reference.  Let us consider an event $A$ at position $P$ in space, which in turn is represented by the set of coordinates $\{ x^i_P \}$\footnote{Greek indices run from ($0\ldots3$) and lowercase Latin indices run from ($1\ldots3$).}.  In our new formulation of physical laws, how do we define the time that the event $A$ occurs?  Without distant simultaneity we can only define time with local clocks and we cannot compare the times of distant clocks.   Therefore, the time that event $A$ occurs depends on the location of the observer.  There is no objective time that can be attributed to the event $A$ throughout space.  An observer collocated with $A$ would define the time of event $A$ as $t_P^A$, which is the local time that is simultaneous with the event $A$.  Note that simultaneity can indeed be defined with a single clock for events collocated with that clock.  An observer not collocated with $A$, say, an observer at a point $Q$ in space, $\{ x^i_Q \}$, will define the time that $A$ occurs as $t_Q^A$, which is the time he receives a light signal that travels from  $\{ x^i_P \}$ to $\{ x^i_Q \}$ that conveys the information that the event $A$ is occurring.  We are tempted and conditioned to retrace light's propagation and compare $t_P^A$ with $t_Q^A$.  However, since nature prevents us from defining distant simultaneity unambiguously, we assert that no further analysis can be achieved.   In other words, we cannot go further to reduce the observed time to be related to a global time with an associated duration that can be attributed to light propagation.  Each observer will assign a time to an event based on the time light is received that communicates that event, and no comparison of the times recorded by separated clocks can be accomplished.  This time for each observer is the \textit{real} time attributed to the event.

We now have an operational definition of time that is closer to reality than Einstein's global time defined throughout an inertial frame.  Our operational definition of time consists of an infinite set of non-synchronized clocks situated throughout space with local observers describing the temporal evolution of distant events using local clocks and information obtained via electromagnetic communication.  Such a definition of time certainly does not provide the most simple formulation of physical laws, nevertheless, it avoids the introduction of the unphysical and unmeasurable one-way speed of light into the mathematical formulation.  Let us go further.  Consider a scalar field, $\phi(x^i)$, defined throughout space.  How do we define the temporal evolution of our field?  As before, we consider an infinite set of non-synchronized clocks, $t_N$, where $N$ refers to the collection of points that define their locations in space, $\{ x^i_N \}$.  Since there is no way to synchronize clocks unambiguously throughout space, one can only define the values of the field at times according to local clocks.  Therefore, consider an observer situated at position $\{ x^i_P \}$ in space. This observer possesses a clock with time values denoted by the quantity $t_P$.  The values of the field $\phi(x^i)$ as a function of time are then the values obtained at times $t_P$ that are communicated from each point in space $\{ x^i \}$ to the observer at $\{ x^i_P \}$ .  The observer at $P$ will collect the totality of values received at each time $t_P$ and construct the temporal evolution of the field $\phi(x^i, t_P)$.  Therefore, a field evaluated at a time $t_P$ according to the local clock at $P$ is the value of the field transmitted from each point in space to the location of the clock and received at the time $t_P$.  We are tempted and conditioned to interpret this as signals sent from each point in space at different times, but we must remind ourselves again that there is no way to unambiguously synchronize clocks throughout space, and therefore we have abandoned such concepts.  The values of the field at each local time will depend on the location of the chosen observer in space and there is no way to compare these values because unambiguous synchronization of distant clocks is not permitted by nature.

One-way velocity can be defined in this description of nature, although it becomes dependent on the distance of the observer from the beginning and end points of the propagation path.  Let us consider an object traveling from point $P$ to point $Q$ with an observer situated at point $R$.  The moment of departure from $P$ according to the observer at $R$ will be the time light is received at $R$ and sent from $P$ at the time simultaneous (according to clocks at $P$) with departure from $P$.  Similarly, the arrival time at $Q$ according to the observer at $R$ will be the time light is received at $R$ that was sent from $Q$ at the time simultaneous (according to clocks at $Q$) with arrival at $Q$.  Velocity can then be defined in the usual manner as the total distance separating $P$ and $Q$ divided by the total time of propagation according to the observer at $R$.  Note that this definition also applies to light propagation.  For example, an observer situated at point $P$ will define the `one-way velocity' of light as $c/2$ and an observer situated at point $Q$ will define the `one-way velocity' of light as infinite.  Interestingly, the observer at $P$ will define the two-way speed of light as $c$ even though the light propagation is over the same distance as the one-way speed calculation.

\section{\label{sec:restricted}Restricted Synchrony Invariance}
Consider an observer at an arbitrary point, $P$, in space with the associated local time coordinate $t_P$.  According to our proposed reformulation of physical laws, this observer will describe events throughout space using the set of coordinates $\{x^i, t_P\}$.  We emphasize that the time the observer at $P$ attributes to an event at an arbitrary point in space, $x^i$, is defined as the time the observer at $P$ receives the information of that event via light signals propagating radially from $x^i$ to $x^i_P$.  Of course, we can consider other forms of communication besides electromagnetic propagation to define the times of distant events and the laws of physics must remain invariant under transformations to frames that use these different forms of communication, as long as the homogeneity of space is preserved.  Therefore, the laws of physics must be invariant under the following transformation of local time:
\begin{equation}
\label{eq:synchronization_transformations_2}
t_P^{\prime} = t_P +  b^{\star}\frac{r}{c},
\end{equation} 
where $r$ is the distance between the event and the observer, $b^{\star}$ is an arbitrary constant, and $c$ is the experimentally-measured two-way speed of light.  This time transformation shifts the readings of the clock at $P$ to change the time attributed to events at a given distance $r$ from the observer to an earlier or later time, depending on the sign of $b^{\star}$.  Identifying electromagnetic propagation as the fastest form of communication will fix the sign of $b^{\star}$.  Note that transformation (\ref{eq:synchronization_transformations_2}) does not attribute a one-way speed of light (in the conventional sense) to light propagation, but introduces a shift in the attributed time that results from employing a different means of communication for all events in space.  The homogeneity of space requires the same means of communication to be utilized throughout space, otherwise the laws of physics would not be invariant.  Transformation (\ref{eq:synchronization_transformations_2}) is to be contrasted with the synchrony transformation of the conventional formulation:
\begin{equation}
\label{eq:synchronization_transformations}
t^{\prime} = t +  b_i \frac{x^{i}}{c},
\end{equation} 
where $b_i$ are arbitrary constants and $t$ is the global time defined using distant simultaneity.  We see that the invariance principle in our reformulation of physical laws is related to a subset of the synchrony transformations of the conventional formulation.  Note that without the proposed observer-centric formulation of physical laws, we would not have arrived at the restricted group of synchrony transformations (\ref{eq:synchronization_transformations_2}).  

\section{\label{sec:example}Static, Spherically Symmetric Solution}
As we showed previously \cite{bock2015a} using the gauge prescription, the group of synchrony transformations (\ref{eq:synchronization_transformations}) leads to the introduction of a new set of fields, $B^{i}_{\;\;\mu}$.  These new fields transform under synchrony transformations as:
\begin{equation}
\delta B^{i}_{\;\;\mu} = -b^{i}_{\;\;,\mu}.
\end{equation}
to retain invariance of the action when the parameters of the synchrony group become arbitrary functions of the coordinates.
We write the Lagrangian for the free fields as 
\begin{equation}
L_0 =  - \frac{1}{4}F_0,
\end{equation}
where $F_0 = F^{i}_{\;\;\mu\nu}F_{i}^{\;\;\mu\nu}$ and ${F}^{i}_{\;\;\mu\nu}  = B^{i}_{\;\;\mu,\nu} - B^{i}_{\;\;\nu,\mu}$ is calculated from the commutator of b-covariant derivatives.  This produces the following field equations:
\begin{equation}
\label{eq:synchrony_field_equations}
 F_{i\;\;;\nu}^{\;\mu\nu}   = J^{\mu}_{\;\; i},
\end{equation}
where $J^{\mu}_{\;\;i} \equiv -\partial L/\partial B^i_{\;\mu}$ and $L$ is the Lagrangian that is a function of a set of fields.  We see that a local gauge theory of the synchrony group in the absence of gravity possesses three sets of fields, each satisfying Maxwell's equations to lowest order.  This is the dilemma mentioned above regarding the challenge of identifying the electromagnetic field as a gauge field of the synchrony group.  The synchrony group generates three fields that satisfy Maxwell's equations to lowest order, rather than a single field.  Note that local synchrony transformations generate transformations that resemble electromagnetic gauge transformations. 

According to our discussion above, we can now reduce these equations to the equations of a single field, $B_\mu$, such that:
\begin{equation}
\label{eq:synchrony_field_equations2}
 F_{\;\;;\nu}^{\mu\nu}   = J^{\mu},
\end{equation}
where ${F}_{\mu\nu}  = B_{\mu,\nu} - B_{\nu,\mu}$ and we identify $B_\mu$ with the electromagnetic potentials, $\phi_\mu$:
\begin{equation}
\label{eq:potential_def}
B_\mu = \alpha\phi_\mu,
\end{equation}
where $\alpha$ is a constant.  For comparison with Maxwell's equations, we consider the equations for a static, spherically symmetric solution, setting $\phi\equiv\phi_r$:
\begin{equation}
\frac{\partial^2 \phi}{\partial r^2} + \frac{\partial \phi}{\partial r}\left(\frac{2}{r}-\alpha\phi      \right) =0.
\end{equation}
Substituting $\psi\equiv\frac{\partial \phi}{\partial r}$, we obtain the following first-order equations:
\begin{eqnarray}
\frac{\partial \phi}{\partial r} &=& \psi \nonumber \\
\frac{\partial \psi}{\partial r} &=&-\psi\left( \frac{2}{r} -\alpha\phi      \right),
\end{eqnarray}
which yields the following implicit equation for the potential:
\begin{equation}
\phi=\int \frac{C}{r^2}\left[e^{\alpha\int\phi dr}\right]dr,
\end{equation}
where $C$ is an arbitrary constant.  We see that setting $\alpha=0$ produces Coulomb's law so that experimentally measured deviations from Coulomb's law can serve to quantify the constant $\alpha$ in Equation (\ref{eq:potential_def}).

\section{\label{sec:discussion}Discussion}
The traditional formulation of physical laws, based on Einstein's theory of special relativity, utilizes distant simultaneity, which is an artificial construct that has been introduced for the expedience of calculations.  Since it is impossible to synchronize spatially separated clocks, we propose that physical laws must be formulated without a global time variable in a given inertial frame of reference.  According to our proposal, the time an observer attributes to a distant event is the time the observer receives information about that event.  Electromagnetic communication can encode information of distant events and due to its unique status is the preferred form of communication for the formulation of physical laws.   However, other forms of communication can be used and the laws of physics must remain invariant under changes in the means of communication chosen.  This observation leads to a symmetry principle that is related to a subset of the synchrony group.  The field that results when gauging this symmetry group produces a field that is identified with the electromagnetic field.

Our reformulation of physical laws based on the removal of distant simultaneity requires that space-time observations be made from a specific observer that utilizes light propagation to obtain information from distant locations and as such bears a striking resemblance to quantum theory.  Indeed, quantum theory has demonstrated the importance of the observer.  Consequently, we propose that the conflict between classical theory and quantum theory resides in the former's adoption of distant simultaneity, which results in an artificial reference frame.  Classical theory represents the limit in which distant simultaneity can be achieved throughout space.  Quantum theory, on the other hand, results from the reality that we cannot synchronize distant clocks yet we use frames of reference with distant simultaneity for our description of physical laws.  The statistical interpretation of nature forced by quantum theory can be understood conceptually then as a result of utilizing a set of perfectly synchronized clocks to formulate laws, which conflicts with the reality that distant clocks cannot be synchronized.  Indeed, time has a preferred status in quantum theory as a coordinate, rather than an operator.  Our current analysis suggests that a reality can indeed be attributed to distant events, even if not observed by a specific observer, but the description of events is intimately connected with the chosen observer who relies on light communication to convey the information from distant locations.  Such a viewpoint resolves Schrodinger's cat paradox, for example, and points towards a path of reconciliation between classical and quantum physics.  

An inertial frame with distant simultaneity can be viewed as complementary (or dual) to an inertial frame with an infinite number of non-synchronized clocks.  These two treatments of time are mutually exclusive and are likely the source of the wave-particle duality of quantum theory.  Consider a frame, $R$, that consists of an infinite set of perfectly synchronized clocks with coordinate time $t$.  We also consider an inertial frame, $S$, that consists of an infinite number of non-synchronized clocks with local time $t_N$, where $N$ denotes the position in space, $x^i_N$.  The observers with clocks of frame $S$ are not permitted to communicate with the observers with clocks of frame $R$.  A measure of non-synchronization can be introduced by calculating the variance according to:
\begin{equation}
{\Delta t}^2=\frac{1}{N}\sum_i^N \left(t_i-t     \right)^2,
\end{equation} 
calculated at any arbitrary time defined by the synchronized frame, and summed over all the locations (denoted by $i$ in the sum) in space.  Given that we have identified the classical and quantum limits as $\Delta t\rightarrow0$ and $\Delta t\rightarrow\infty$, respectively, we make the following identification:
\begin{equation}
\Delta t = \frac{h}{m_0 c^2},
\end{equation}
where $m_0$ is the rest mass and $h$ is Planck's constant.  In other words, infinite mass represents the limit that perfect synchronization can be achieved by clocks in a reference frame and vanishing mass denotes the opposite limit of non-synchronized clocks with infinite spread in their readings.  Mass can then be interpreted as a measure of the non-synchronization of clocks of the underlying frame.  The observed spectra of fundamental particles, such as the leptons, suggests that in the limit of infinite spread in the measure of non-synchronization, nature prefers specified, discrete values. Furthermore, the observed spectra should be able to be traced to the statistics of the allowed sets of non-synchronized clocks.  According to this interpretation, photons represent the limit of non-synchronized clocks with infinite spread in their readings.

It is important to point out that we assumed infinite bandwidth in communication for conveying the information of distant events throughout space to a local observer.  In other words, we have assumed that the electromagnetic signals (or another chosen form of communication) can carry an infinite amount of information from each distant point in space to the location of the clock chosen for the formulation of physical laws.  This assumption may be problematic from an operational standpoint and may require further consideration.

\bibliographystyle{unsrt}
\bibliography{paper}

\end{document}